\documentclass[amssymb,prb,twocolumn,showpacs,superscriptaddress]{revtex4}
\usepackage{epsfig}
\usepackage{dcolumn}
\usepackage{amsmath,amssymb}
\begin{document}

\title{Majorana modes and complex band structure of quantum wires} 

\author{Lloren\c{c} Serra}
\affiliation{Institut de F\'{\i}sica Interdisciplin\`aria i de Sistemes Complexos
IFISC (CSIC-UIB), E-07122 Palma de Mallorca, Spain}
\affiliation{Departament de F\'{\i}sica,
Universitat de les Illes Balears, E-07122 Palma de Mallorca, Spain}

\date{October 17, 2012}

\begin{abstract}
We describe Majorana edge states of a semi-infinite wire using the
complex band structure approach. In this method the edge state
at a given energy is built as 
a superposition of evanescent waves. It is shown that the superposition 
can not always
satisfy the required boundary condition, thus restricting the existence
of edge modes. We discuss purely 1D and 2D systems,
focussing in the latter case on the effect of the Rashba 
mixing term.
 \end{abstract}

\maketitle

\section{Introduction}

The realization of Majorana states as  specific excitations of 
semiconductor quantum 
wires and of other condensed matter systems 
has attracted much interest 
recently.\cite{ali12,fu08,akh09,tan09,law09,lut10,ore11,fle10,pot10,akh10,gan11,
egg10,lut11,sta11,zaz11,wim11,zit11,gol11,li12,kli120,kli12,bee12,
mou12,den12,rok12,das12,sas11,wil12}
Experimental evidences of these peculiar states in quantum wires
and topological insulators have been presented in Refs.\ [\onlinecite{mou12,den12,rok12,das12}]
and [\onlinecite{sas11,wil12}], respectively.
To assess these evidences, theoretical works 
are presently focussing on a deeper clarification of the 
physical properties of Majorana states, including analysis of stability limitations
and robustness against disorder and other 
perturbations.\cite{lob12,tew12,lim12,lim12b,pik12,coo12,rain12}

As most characteristic properties, a Majorana state is 
a zero-energy mode, thus degenerate with the system ground state,
that is localized on the system edges or interfaces. Additionally, an energy gap
with neighboring states protects the Majorana modes against decoherence.
In semiconductor wires, the subject of this work, such zero-energy modes can
be induced by the combined action of the following three mechanisms: 
superconductivity, Rashba spin-orbit coupling and Zeeman magnetic field.
Superconductivity introduces particle-hole duality, with 
excitations lying symmetrically at positive and negative energies
with respect to the chemical potential. Rashba
spin-orbit coupling introduces helicity by associating motion in space
with spin. Thirdly, the Zeeman coupling with an applied magnetic field
can be viewed as a tunable interaction 
with which the system can be placed in different regimes.

The edge character of the Majorana modes suggests a link
with the physics of surfaces. Indeed, the similarities with the Schockley
states of finite crystals \cite{sho39} have been pointed out in Ref.\ [\onlinecite{bee12}]. 
We explore in this work another link  
with surface physics, namely the description of 
Majorana modes using the complex band structure of the system.
The complex band structure of crystals (or periodic systems in general) is a well established theoretical formalism 
allowing the generalization from bulk propagating states to states that are localized
around defects or interfaces of the crystal.\cite{tom02} The underlying 
idea of the generalization is the following: while usual propagating states 
are described by sustained
waves with a real wave number $k$, there are 
other solutions with complex wave numbers representing evanescent waves whose amplitude 
decreases in a given direction (increases in the reversed one).

Waves with complex $k$ cannot be physically realized
for infinitely long 
distances from a given point
in both 
directions 
due to the divergent behavior in one of them; but they can in a finite 
domain, or even in a semi-infinite one provided the infinite distance is in the  
direction of decaying amplitude. The latter situation is precisely the
one we expect for edge Majorana modes. While for propagating states
the analysis is usually done in terms of the band structure $\{E_\alpha(k)\}$,
where $\alpha$ is the band index and $k$ is an arbitrary real wave number, 
the complex band structure is presented as $\{k_\alpha(E)\}$, where
for any given energy $E$ there is a set of allowed complex wave numbers ${k_\alpha}$.
Of course, if the energy does not allow propagating states 
all $k_\alpha$ will have an imaginary part. In general, however, 
propagating and evanescent modes may coexist.

In this work we determine numerically the complex band structure of a semiconductor 
quantum wire with the above mentioned interactions. Previously,\cite{ser07}
we studied the complex band structure 
in absence of superconductivity and Zeeman field. 
Our present numerical approach to Majorana physics complements the analytical 
methods used in Ref.\ [\onlinecite{kli12}]. We suggest robust numerical algorithms
to obtain the complex band structure of 1D and 2D quantum wires 
in general, 
discussing
the conditions for the existence of Majorana modes in the semi-infinite
system. 
In 2D we focus on the role of the Rashba mixing, giving
evidence that the Rashba mixing mechanism hinders the coexistence of two
Majorana modes. This is in agreement with the findings of numerical 
diagonalizations for the finite system.\cite{lim12} 
Our results correspond to a continuum Hamiltonian of 
quantum wires
and, from this point of view, they also complement approaches based
on tight binding models.\cite{pot10,sta11}

The analysis in terms of the complex band structure gives the precise 
conditions for the existence of zero-energy edge modes, as
an alternative to the analysis based on propagating bands.
In particular, it clarifies why the topological phase transition 
from absence to presence of a Majorana mode
is signaled by the vanishing and reopening of the gap for 
propagating states with $k=0$, as discussed in Ref.\ [\onlinecite{ore11}].
Sections II and III are devoted to the 1D and 2D cases, respectively. 
The first one is meant as a simple yet illustrative case of the 
more realistic situation including transverse degrees of freedom.
Being well understood,\cite{ali12,fu08,akh09,tan09,law09,lut10,ore11,fle10,pot10,akh10,gan11,
egg10,lut11,sta11,zaz11,wim11,zit11,gol11,li12,kli120,kli12,bee12} 
the 1D system serves as a benchmark of the 
complex band structure method.

\section{The 1D case}

\subsection{Hamiltonian and complex band structure}

Let us first consider a purely 1D model as in Ref.\ [\onlinecite{kli12}], 
with motion constrained to be on the $x$ axis. The Hamiltonian reads in this case
\begin{eqnarray}
\label{eq1}
{\cal H}_{1D} &=& \left(
\frac{p_x^2}{2m}-\mu
\right)\tau_z + \Delta_B \vec{\sigma}\cdot\hat{n}+\Delta_0\tau_x\nonumber\\
&+& \frac{\alpha}{\hbar} p_x \sigma_y \tau_z\; ,
\end{eqnarray}
where the Pauli operators for spin and isospin (in electron-hole space)  
are represented by the $\sigma$'s and
$\tau$'s, respectively. The successive energy contributions to Eq.\ (\ref{eq1})
are: kinetic, Zeeman (depending on parameter $\Delta_B$), pairing
(parameter $\Delta_0$) and Rashba spin-orbit coupling (parameter $\alpha$).
In addition, Eq.\ (\ref{eq1}) includes the chemical potential ($\mu$)
and the magnetic field orientation ($\hat{n}$). 

The system exhibits 
very different physics depending on the field orientation. 
In 1D the Rashba term in Eq.\ (\ref{eq1}) is defining a preferred spin
direction ($y$) so that, in abscence of any orbital effects, the physics depends 
on the field orientation relative to this Rashba-induced direction.  
Since field orientations along $x$ and $z$ are equivalent we will then consider a field lying
on the $xy$ plane, with an azimuthal angle $\phi_B$ ($0\le\phi_B\le 2\pi$).
The two limiting cases are the parallel ($\hat{n}=\hat{x}$) and 
transverse ($\hat{n}=\hat{y}$) orientations with respect to the wire.

The band structure of the infinite system is obtained from Schr\"odinger equation
\begin{equation}
{\cal H}_{1D} \Psi(x,\eta_\sigma,\eta_\tau) =
E \Psi(x,\eta_\sigma,\eta_\tau)\;,
\end{equation} 
where the wave function variables are the space coordinate $x\in (-\infty,+\infty)$, 
the spin 
$\eta_\sigma\in\{\uparrow,\downarrow\}$
and the isospin 
$\eta_\tau\in\{\Uparrow,\Downarrow\}$.
Spin and isospin basis  are taken in $z$ orientation which means, 
for instance, that $|\Psi(x,\uparrow,\Uparrow)|^2$ is the probability of having
at $x$ a quasiparticle  with spin pointing along $+z$
($\uparrow$) of electron type ($\Uparrow$).

Introducing a state wave number $k$ we now assume the following
expansion
\begin{equation}
\label{eq3}
\Psi(x,\eta_\sigma,\eta_\tau)
=
\sum_{s_\sigma s_\tau}{
\psi^{(k)}_{ s_\sigma s_\tau}\,\,
e^{ikx}\,
\chi_{s_\sigma}(\eta_\sigma)\,
\chi_{s_\tau}(\eta_\tau)
}\; ,
\end{equation}
where the quantum numbers are $s_\sigma=\pm$, $s_\tau=\pm$ and the 
spin/isospin two-component states fulfill
\begin{eqnarray}
\vec{\sigma}\cdot\hat{n}\, \chi_{s_\sigma}(\eta_\sigma) &=& s_\sigma\, \chi_{s_\sigma}(\eta_\sigma)\; ,\\
\tau_z\, \chi_{s_\tau}(\eta_\tau) &=& s_\tau\, \chi_{s_\tau}(\eta_\tau)\; .
\end{eqnarray}
The amplitudes $\rule{0cm}{0.7cm}$ $\psi^{(k)}_{s_\sigma s_\tau}$ in Eq.\ (\ref{eq3}) can be viewed as a set of four complex numbers
representing the state for a given wave number $k$.   
Substituting Eq.\ (\ref{eq3}) in Eq.\ (\ref{eq1}) we easily get the equations for the state
complex amplitudes $\psi_{s_\sigma s_\tau}$ (we drop label $k$  for convenience) 
\begin{eqnarray}
\label{eq6}
&& 
\!\!\!\!\!\!\!\!
\left[
\left(
\displaystyle\frac{\hbar^2 k^2}{2m}
-\mu
\right) s_\tau 
+  \Delta_B s_\sigma
+ \alpha k\, s_\sigma s_\tau\sin\phi_B
-E
\right] \psi_{s_\sigma s_\tau} \nonumber\\
&& \hspace*{1cm} + \Delta_0\, \psi_{s_\sigma \overline{s_\tau}}
+ i \alpha k\, s_\sigma s_\tau\cos\phi_B\,
\psi_{\overline{s_\sigma} s_\tau}=0\; ,
\end{eqnarray}
where $\overline{s}=-s$. Equation (\ref{eq6}) is an algebraic (matrix) equation
with nondiagonal couplings of $\psi_{s_\sigma s_\tau}$ to
 $\psi_{s_\sigma \overline{s_\tau}}$
through pairing
and to
$\psi_{\overline{s_\sigma}s_\tau}$ through the Rashba interaction.

It is important to realize that Eq.\ (\ref{eq6}) has solution only for restricted 
values of the wave number $k$. An allowed $k$ which is purely real describes a propagating state 
while  a complex $k$ with nonvanishing imaginary part describes an evanescent state.
Finding the physically allowed $k$'s at a given energy $E$, even numerically, is a non trivial task.
Notice, for instance, that the matrix in Eq.\ (\ref{eq6}) is non Hermitian for an arbitrary 
complex $k$. While many numerical methods can deal with Hermitian matrices, the methods
for non Hermitian operators are scarce. An approach based on the determination of the roots of 
a high-degree polynomial is possible in 1D.\cite{note} We follow, however, an alternative 
method allowing an extension to higher dimensions.  

We have followed the idea suggested in Ref.\ [\onlinecite{ser07}] to find the allowed $k$'s in a 
robust and numerically stable way.
We arbitrarily choose a particular  spin and isospin $(s,t)$ and solve the following linear 
system
\begin{equation}
\label{eq7}
\left\{
\begin{array}{lcl}
\psi_{s_\sigma s_\tau}=1 &\quad& {\rm if\ } (s_\sigma,s_\tau)=(s,t)\; ,\\
{\rm [ Eq.\ (\ref{eq6})] }&\quad& {\rm if\ } (s_\sigma,s_\tau)\ne(s,t)\; .
\end{array}
\right.
\end{equation}
We have found that Eq.\ (\ref{eq7}) is well behaved numerically and always has a solution for any 
wave number $k$. In order to find the physically acceptable $k$'s we subsequently  look for the 
zeros of the following function obtained from the solution of Eq.\ (\ref{eq7}),
\begin{equation}
\label{eq8}
{\cal F}_{1D}(k)=\left| \rule{0cm}{0.5cm}{\rm l.h.s.\ of\ Eq.\ (\ref{eq6})\ for\ } (s_\sigma,s_\tau)=(s,t)\right|^2\; .
\end{equation}
Notice that the state amplitudes and wave numbers  fulfilling ${\cal F}_{1D}(k)=0$ from Eqs.\ (\ref{eq7})
and (\ref{eq8}) are also solutions of Eq.\ (\ref{eq6}) and, therefore, they are the physical
solutions giving the complex band structure of the system.

\begin{figure}[t]
\centerline{
\epsfig{file=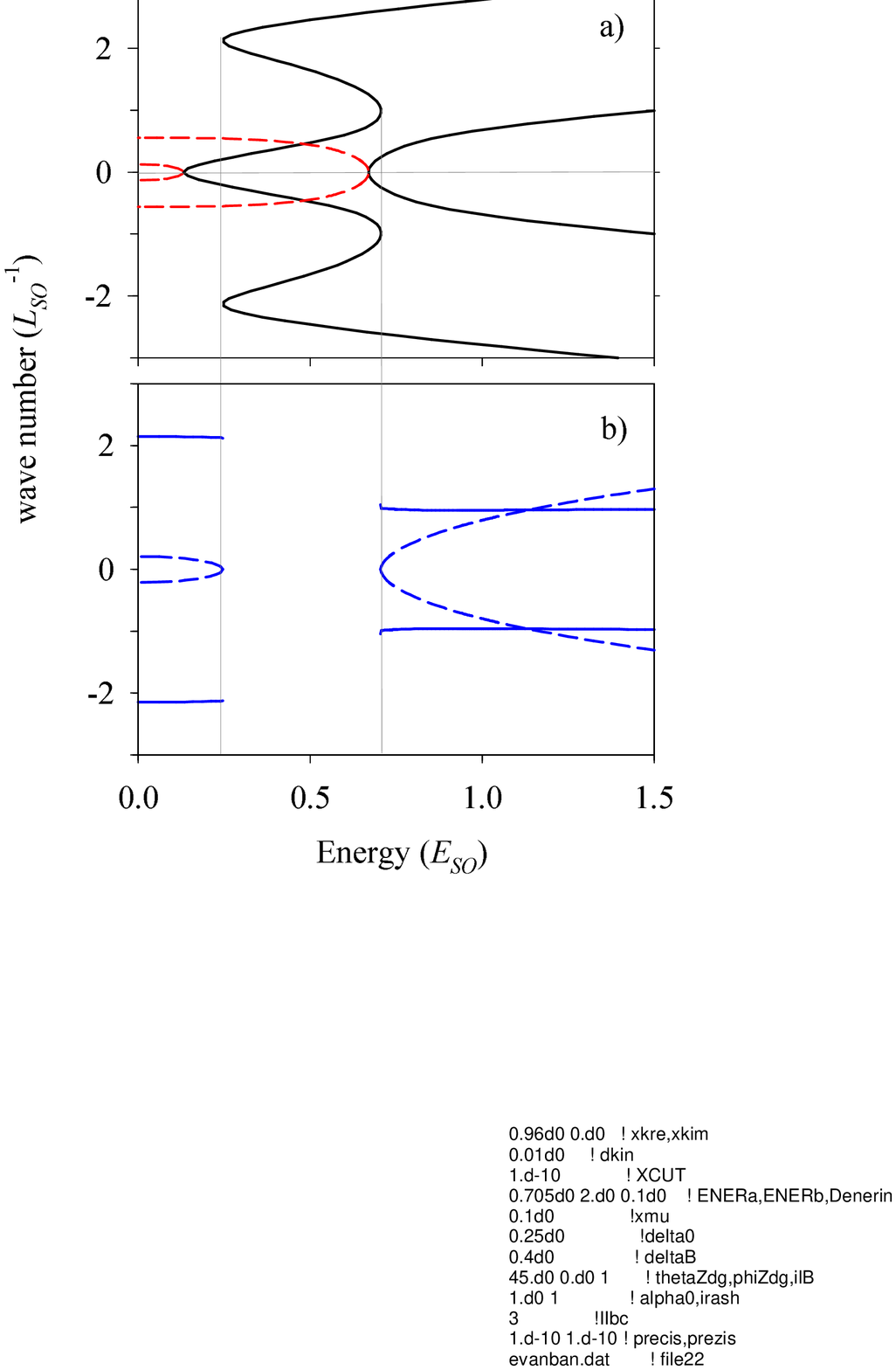,width=0.4\textwidth,clip}
}
\caption{(Color online)
Band structure as a function of energy for $\Delta_B=0.4 E_{SO}$,
$\Delta_0=0.25 E_{SO}$, $\mu=0.1 E_{SO}$, $\phi_B=\pi/4$. Solid and dashed 
lines correspond to $\Re(k)$ and $\Im(k)$, respectively. 
Panel a) shows modes having only one nonvanishing component
of the wave number:
solid lines are for propagating modes with $\Im(k)=0$ while
dashed lines are for evanescent modes with $\Re(k)=0$. Panel
b) shows the complex $k$'s for evanescent modes with both real
and imaginary parts.
The units are defined in Eqs.\ (\ref{eq9n}) and (\ref{eq10n}).
Thin gray lines are guides helping the eye to correlate the 
position of the band extrema.
} 
\label{fig1}
\end{figure}

As an illustrative example, 
Fig.\ \ref{fig1} shows a typical complex band structure. 
Notice that, contrary to common band representations, we 
choose the energy as horizontal axis since, within our approach,
$E$ is given and the wave numbers are inferred.
We only display
the positive energy solutions, the negative
ones corresponding  to the specular image with respect
to $E=0$.
As in Ref.\ [\onlinecite{ore11}] we set our energy units using the 
Rashba interaction as a reference.  Specifically, our
length and energy units are
\begin{eqnarray}
\label{eq9n}
L_{SO}&=& \frac{\hbar^2}{\alpha m}\;,\\
\label{eq10n}
E_{SO}&=& \frac{\alpha^2 m}{\hbar^2}\; .
\end{eqnarray}

The propagating states in Fig.\ $\ref{fig1}$a agree with 
those already discussed in Refs.\ [\onlinecite{ore11,kli12}].
The evanescent states provide, however, a novel view. We distinguish
two types of evanescent modes: those 
shown in Fig.\ \ref{fig1}a
have a purely imaginary
wave number, while those
with wave numbers having both real and imaginary parts  
are shown in panel b). For a given energy, wave numbers
always come in sets $\{k,k^*,-k,-k^*\}$ or, equivalently,
as $\pm|\Re(k)|\pm i|\Im(k)|$, 
where $\Re$ and $\Im$ stand for real and imaginary parts, respectively.\cite{ser07}
That is, wave numbers come 
in quartets when both real and imaginary parts are non vanishing (panel b)
and in pairs when the wave number is either purely real or purely imaginary (panel a).   

Figure \ref{fig1} shows 
that propagating and evanescent bands can be smoothly connected
by the extrema points. In panel a) this connection is 
clearly seen as horizontal parabolas
of opposite curvatures, showing how a purely imaginary 
wave number transforms into a purely real one as the energy increases.
In panel b), $\Re(k)$ of an evanescent state connects with the 
extremum of the corresponding propagating band, while
the imaginary part disappears when entering the propagating
band sector. For a given energy, there are always 8 modes, as
corresponds to 
a quadratic kinetic energy and 
the $4\times 4$ matrix in Eq.\ (\ref{eq6}).

Summarizing this section, we stress that the
suggested approach is a robust numerical algorithm allowing the determination 
of the complex band structure for an arbitrary set of parameters. 
We will focus next on the description of edge states, in particular
zero energy ones representing Majorana modes.

\subsection{Edge states}

The complex band structure contains the necessary information on all
possible eigenstates. In particular, localized states for which the wave function 
fades away of some defect or interface are, in general, superpositions of evanescent
states. Let us consider a semi-infinite 1D system restricted to $x\ge 0$. A priori, 
we expect the abrupt edge at $x=0$ to allow the formation of localized states 
vanishing at $x=0$ and decaying towards $x\to\infty$. Such a state has to be a superposition 
of evanescent waves having $\Im(k)>0$.
Let us assume that the set of such wave numbers is $\{\tilde k\}$, 
containing $\tilde N$ modes
for a given energy $E$.
The boundary condition an edge state has to fulfill is then 
\begin{equation}
\label{eq9}
\sum_{k\in\{\tilde k\}}
C_{k}\, \psi_{s_\sigma s_\tau}^{(k)}=0\; ,
\end{equation}
where the $C_k$'s are complex numbers and the $\psi_{s_\sigma s_\tau}^{(k)}$'s  
are the state amplitudes defined in the preceding subsection.

Equation (\ref{eq9}) is a set of four linear equations, for $(s_\sigma s_\tau)=(++,+-,-+,--)$,
and $\tilde N$ unknown $C_k$'s. We can immediately 
conclude that if $\tilde N < 4$ no edge state is possible since there are less 
unknowns than equations in Eq. (\ref{eq9}). That is, if there are less than 4 evanescent 
modes with $\Im(k)>0$ only the 
trivial solution $C_k=0$ is possible, i.e., no physical 
solution is present. On the contrary, if $\tilde N>4$ it will always be possible to 
find a set of nonvanishing $C_k's$. If $\tilde N=4$ the nontrivial solution will be 
possible if the system matrix is singular (zero determinant). In that case the 
condition for the existence of an edge state is 
\begin{equation}
\label{eq12}
\det\{\psi_{s_\sigma s_\tau}^{(k)}\}=0\; .
\end{equation}

\begin{figure}[t]
\centerline{
\epsfig{file=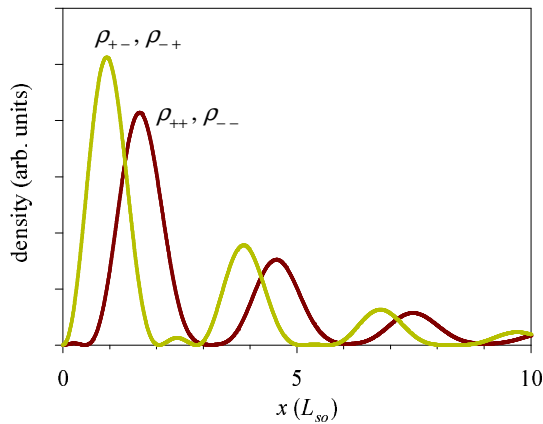,width=0.4\textwidth,clip}
}
\caption{(Color online)
Densities of the Majorana mode (edge state at zero energy)
corresponding to the 
complex band structure of Fig.\ \ref{fig1}.
The definition of $\rho_{s_\sigma,s_\tau}(x)$
is given in Eq.\ (\ref{eq14n}). 
} 
\label{fig2}
\end{figure}

For instance, in the case of Fig.\ \ref{fig1} the requirement
of at least 4 evanescent modes restricts the 
possibility of an edge mode to energies $E<0.1 E_{SO}$. We have 
also computed the determinant of Eq.\ (\ref{eq12}) finding that 
it vanishes only for $E=0$, while it grows linearly for $E>0$. 
Therefore, in this
example a zero energy edge state is formed. This is a Majorana 
mode for which
Fig.\ \ref{fig2} displays the corresponding density profile, 
obtained as
\begin{eqnarray}
\rho(x) &\equiv& 
\sum_{s_\sigma s_\tau}{\rho_{s_\sigma s_\tau}(x)}\; , \\
\nonumber\\
\label{eq14n}
\rho_{s_\sigma s_\tau}(x) &=&
\sum_{kk'}{
C_{k'}^* C_{k}\,
\psi^{(k')*}_{s_\sigma s_\tau}
\psi^{(k)}_{s_\sigma s_\tau}
e^{i(k-{k'}^*)x}
}
\; .
\end{eqnarray}
As expected, the densities vanish at $x=0$ and decay with 
damped oscillations for increasing $x$. 
This behavior is in good agreement with previous 
results.\cite{kli12} Notice also that 
because of symmetry the densities 
remain invariant after spin-isospin inversion.

\begin{figure}[t]
\centerline{
\epsfig{file=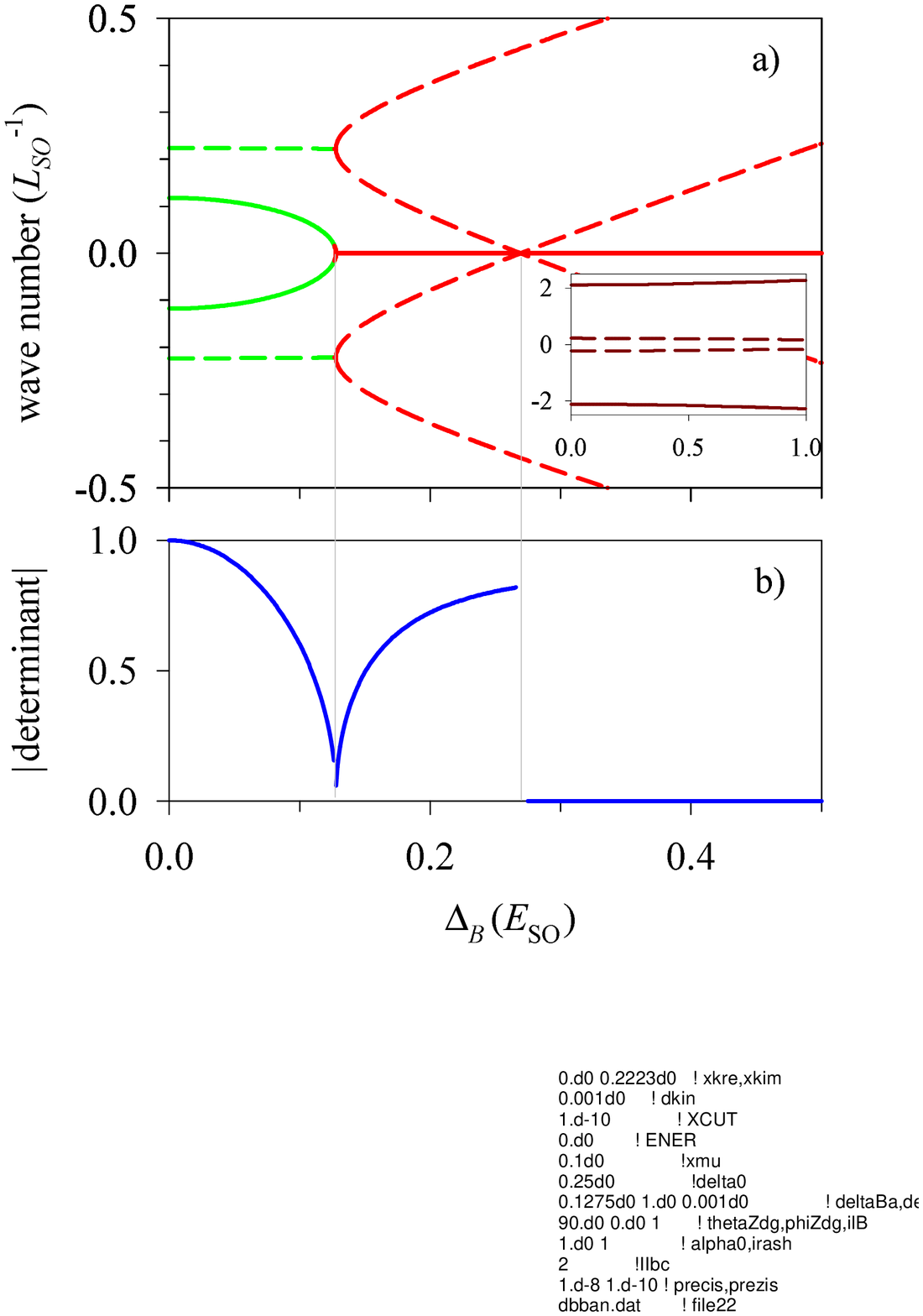,width=0.4\textwidth,clip}
}
\caption{(Color online)
a) Complex band structure as a function of $\Delta_B$ for fixed $E=0$, $\Delta_0=0.25 E_{SO}$,
$\mu=0.1E_{SO}$, $\phi_B=0$. As in Fig.\ \ref{fig1}, solid and dashed lines display real and
imaginary parts of the wave number. The inset shows 4 modes with a large value of $\Re(k)$.
b) Absolute value of the determinant Eq.\ (\ref{eq12}). A vanishing determinant is the condition
for the existence of an edge state.   
} 
\label{fig3}
\end{figure}

The above results can be taken as examples of the method and its 
capabilities and we now address the question of how to rationalize
the existence of topological phases. It was advanced by Oreg {\em et al.}
\cite{ore11}
that the signature of an emergent topological phase is the closing and 
reopening of the 
$k=0$ gap in the band structure of propagating states
as one Hamiltonian parameter is varied. 
Increasing the Zeeman
energy in a parallel field geometry the critical 
parameter for the topological transition is
\begin{equation}
\label{eq15n}
\Delta_B^{(c)}=\sqrt{\Delta_0^2+\mu^2}\; ,
\end{equation}
such that zero-energy edge (Majorana) modes are present
only when $\Delta_B>\Delta_B^{(c)}$. In the transverse field
geometry, no zero energy modes are ever present. Let us analyze this 
scenario from the evanescent band structure point of view.

Figure \ref{fig3} shows the evolution of the wave numbers at zero energy 
as a function of the Zeeman parameter. The field orientation is along $x$ 
and, in this case, we always find 8 evanescent modes for any $\Delta_B$.
This means that there is no propagating mode for this energy. Notice 
that four evanescent modes with a rather structureless behavior
and having a large real part of the wave number
are shown in the inset. 
In panel b) we display the absolute value of the determinant
Eq.\ (\ref{eq12}) as a function of $\Delta_B$. The determinant has a discontinuous
behavior: it decays from its maximal value for $\Delta_B=0$; it has
an accidental zero for $\Delta_B\approx 0.13E_{SO}$, when all wave numbers
in Fig.\ \ref{fig3}a
become purely  imaginary; and it consistently vanishes for $\Delta_B>0.27E_{SO}$, 
after the point where $\Im(k)$ vanishes for two of the modes.

\begin{figure}[t]
\centerline{
\epsfig{file=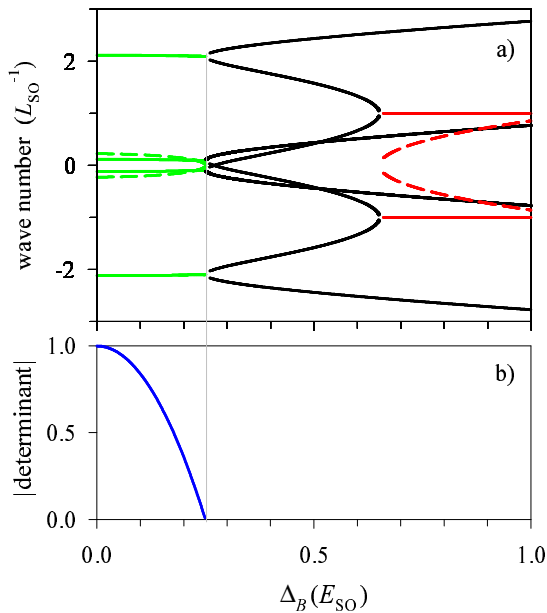,width=0.4\textwidth,clip}
}
\caption{(Color online)
Same as Fig.\ \ref{fig3} for field orientation along $y$ ($\phi_B=\pi/2$).
In panel b) the determinant can only be defined for  $\Delta_B<0.25 E_{SO}$,
when there are 4 evanescent modes with $\Im(k)>0$.} 
\label{fig4}
\end{figure}

The results in Fig.\ \ref{fig3} show
that Majorana modes exist above a critical field ($0.27E_{SO}$ in that case)
which is in agreement with the $\Delta_B^{(c)}$ of Eq.\ (\ref{eq15n}) 
anticipated by Oreg and collaborators.\cite{ore11} Not only this, the analysis 
in terms of evanescent modes also sheds light on the reason why the topological 
transition is signaled by the vanishing of a gap for propagating states. Indeed,
when $\Delta_B=\Delta_B^{(c)}$ in Fig.\ (\ref{fig3}) the complete vanishing of $k$ for 
two of the modes indicates that they loose the evanescent character and become propagating,
but only for a single point in $\Delta_B$ and thus they are not seen in panel a)
as fully developed bands. At this value of $\Delta_B$, therefore, there is no energy gap for 
propagating states. Notice, finally, that Fig.\ \ref{fig3} also shows that the vanishing
of the gap for propagating states occurs for a wave number $k=0$.

When the field is in transverse direction ($y$),
as in Fig.\ \ref{fig4}, 
no Majorana modes are present. Quite remarkably, in this case there are propagating modes
at zero energy (dark solid lines appearing for $\Delta_B>0.25 E_{SO}$).
The condition of at least four evanescent modes 
with $\Im(k)>0$
is fulfilled only when $\Delta_B<0.25E_{SO}$
but, even in this region, no edge state is allowed because of the non vanishing determinant,
as shown in panel b).
 
\section{The 2D case}

We consider now the generalization of the complex band structure approach to include one extra spatial degree of freedom in the transverse direction to the wire. As before, we assume the wire to be infinite along
$x$, but now motion along $y$ is also possible with the restriction $-L_y/2\le y\le L_y/2$. This corresponds
to a hard wall confinement in $y$ within a square well of length $L_y$. We call this a 2D case although
in the literature the name quasi-1D is often used to emphasize that there is confinement
in the  transverse direction while along the longitudinal one motion is free. 

\subsection{Hamiltonian}
The generalization of Hamiltonian (\ref{eq1}) is
\begin{eqnarray}
\label{eq11}
{\cal H}_{2D} &=& \left(
\frac{p_x^2}{2m}+
\frac{p_y^2}{2m}+
V(y)
-\mu
\right)\tau_z + \Delta_B \vec{\sigma}\cdot\hat{n}\nonumber\\
&+& \Delta_0\tau_x + \frac{\alpha}{\hbar} \left(p_x \sigma_y-p_y\sigma_x\right) \tau_z\; ,
\end{eqnarray}
where $V(y)$ is the square well of length $L_y$ mentioned above. Besides $y$ confinement
Hamiltonian (\ref{eq11}) includes the transverse kinetic energy and a $p_y$-dependent Rashba
term that, following a usual convention, we call the Rashba mixing term. The {\em mixing} 
character of this term can be understood from the fact that the $p_y$ operator
couples different square well eigenstates.

The analysis proceeds in a similar way to the preceding 1D case with the important difference that 
now the state amplitudes of Eq.\ (\ref{eq3}) become functions of $y$,
\begin{equation}
\psi_{s_\sigma s_\tau}
\to\psi_{s_\sigma s_\tau}(y)\; .
\end{equation}
Accordingly, the algebraic Eq.\ (\ref{eq6}) becomes a second order differential equation 
due to the dependence on $p_y^2$ and $p_y$
\begin{eqnarray}
\label{eq13}
&& \left[\rule{0cm}{0.6cm}
\left(\frac{p_y^2}{2m}+
\displaystyle\frac{\hbar^2 k^2}{2m}
-\mu
\right) s_\tau 
+  \Delta_B s_\sigma+
\alpha k s_\sigma s_\tau \sin\phi_B
\right.\nonumber\\
&& \hspace*{1cm}\left.\rule{0cm}{0.6cm}
-E
\right] \psi_{s_\sigma s_\tau}(y) 
+ \Delta_0\, \psi_{s_\sigma \overline{s_\tau}}(y)\nonumber\\
&& + \alpha s_\tau 
\left(i k\, s_\sigma \cos\phi_B\,
-\frac{p_y}{\hbar}\right)
\psi_{\overline{s_\sigma} s_\tau}(y)
=0\; .
\end{eqnarray}

Another difference with respect to the 1D case  
is the dependence on the magnetic field orientation.
Now $x$ and $z$ orientations are no longer physically equivalent due to the orbital
currents present in the latter.
However,  in Eq.\ (\ref{eq13}) we have again assumed a magnetic
field in the $xy$ plane with an azimuthal angle $\phi_B$ since,  in this geometry, there are
no orbital effects of the field.

From the numerical point of view solving Eqs.\ (\ref{eq13}) for an arbitrary $k$ is much more demanding 
than the 1D case. We have devised a practical algorithm,\cite{ser07}
introducing a uniform grid in $y$ and an  arbitrary {\em matching point} $y_m$. 
The differential equation
(\ref{eq13}) on all grid points
but $y_m$ is rewritten
using finite difference formulas, with an important
caveat: crossing from left to right of the matching point is avoided
using non centered finite difference formulas. 
Another condition is that the state amplitude vanishes at the edges of the $y$ domain.
At the matching point $y_m$, instead of Eq.\ (\ref{eq13}), we require the conditions
\begin{equation}
\label{eq14}
\begin{array} {rcll}
\psi_{s_\sigma s_\tau}(y_m)&=&1\;,& {\rm if}\; (s_\sigma,s_\tau)=(s,t)\,,\\
\left(\frac{d^{(L)}}{dy}
-
\frac{d^{(R)}}{dy}\right)
\psi_{s_\sigma s_\tau}(y_m) &=&0\;,& {\rm if}\; (s_\sigma,s_\tau)\ne(s,t)\,,
\end{array}
\end{equation}
where $(s,t)$ are a pair of arbitrarily chosen spin and isospin labels.
In Eq.\ (\ref{eq14}) we have used the notation $d^{(L,R)}/dy$ to indicate
derivatives at $y_m$ using only left ($L$) or right (R) neighboring grid points.
Of course, a physically valid wave number must correspond to a continuous first 
derivative of $\psi_{st}(y)$ at the matching point. We thus numerically 
determine $k$ from the 
zeros of the function
\begin{equation}
{\cal F}_{2D}(k) =
\left|
\left(\frac{d^{(L)}}{dy}
-
\frac{d^{(R)}}{dy}\right)
\psi_{st}(y_m) 
\right|^2\; .
\end{equation}   
As in the 1D case, when ${\cal F}_{2D}(k)=0$ is fulfilled the solution from
Eq.\ (\ref{eq14}) is equivalent to that of the full Eq.\ (\ref{eq13}).

The robustness of the suggested numerical algorithms, both for 2D and 1D, 
is seen in the fact that for any complex $k$ the ${\cal F}$ functions can be
computed, avoiding singular matrices or ill-behaved numerical
problems.\cite{tom02} Finding the physically allowed $k$'s is then 
accomplished
scanning the complex $k$ plane to determine the zeros of ${\cal F}$ with the
desired accuracy. We have checked that the solution is not affected by the 
arbitrary choice of state components $(s,t)$ and matching point $y_m$.
Besides, since Eq.\ (\ref{eq13}) involves only one variable ($y$), high numerical
precission can be easily obtained using large enough numbers of grid points 
($N_y\approx 500$) and of points for the finite difference derivatives
(5 to 11 points).

\subsection{2D edge states}

In 2D the boundary condition for an edge state with $x\ge0$ is similar to Eq.\ (\ref{eq9}),
with the state amplitudes changed 
from scalars to $y$-functions,
\begin{equation}
\label{eq16}
\sum_{k\in\{\tilde k\}}
C_{k}\, \psi_{s_\sigma s_\tau}^{(k)}(y)=0\; .
\end{equation}
We immediately notice that now the boundary condition requires an infinite number of evanescent states
$\{\tilde{k}\}$ due to the infinite number of possible values of $y$. In practice, 
a truncation to a restricted set of $\tilde{N}$ evanescent modes has to be done and, accordingly, 
the  condition (\ref{eq16}) can be imposed on $\tilde{N}/4$ values of $y$ (the number of modes
should be a multiple of 4).

Instead of using
specific values of $y$ in Eq.\ (\ref{eq16}),  we have projected Eq.\ (\ref{eq16}) on the
set of $\tilde{N}$ evanescent modes, finding the matrix equation
\begin{equation}
\label{eq17}
\sum_{k\in\{\tilde k\}}
 {\cal M}_{k'k}\,C_{k}=0\; ,
\end{equation}
where
\begin{equation}
\label{eq23}
{\cal M}_{k'k}=
\sum_{s_\sigma s_\tau}\int{dy\,
\psi_{s_\sigma s_\tau }^{(k')*}(y)\,
\psi_{s_\sigma s_\tau }^{(k)}(y)\,
\;.}
\end{equation}
When the number $\tilde{N}$ of evanescent modes is large enough, the zero eigenvalues of
matrix ${\cal M}$ correspond to Majorana edge states.  This Hermitian matrix can be diagonalized numerically with standard methods. For the 1D case of the preceding section the diagonalization
of matrix ${\cal M}$ is equivalent to the approach based on the determinant, Eq.\ (\ref{eq12}), 
since a zero eigenvalue is associated with a vanishing determinant. 
In 2D matrix ${\cal M}$ allows a truncation of the infinite number of evanescent modes.

\begin{figure}[t]
\centerline{
\epsfig{file=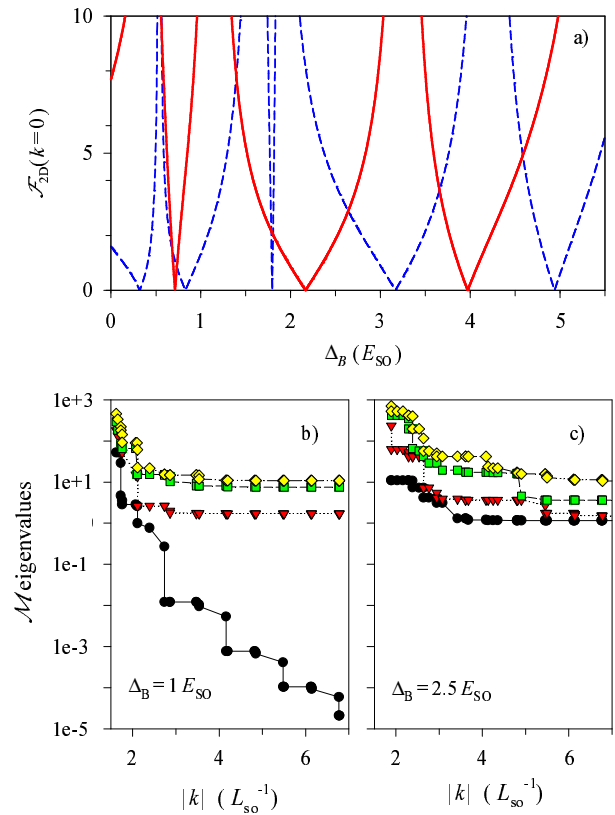,width=0.45\textwidth,clip}
}
\caption{(Color online)
a) Value of the derivative discontinuity ${\cal F}_{2D}$ for $k=0$. 
Solid and dashed lines correspond to presence and absence of the Rashba
mixing term, respectively.
In absence of Rashba mixing (dashes), the nodes signal the 
appearance of successive Majorana modes when increasing $\Delta_B$.
For the case with Rashba mixing, panels
b) and c) show the evolution of the lower eigenvalues of matrix ${\cal M}$
when increasing the number of evanescent modes. The appearance of 
a zero eigenvalue is seen in b).
Parameters:
$\Delta_0=0.25 E_{SO}$, $\mu=0$, $E=0$, $L_y=5L_{SO}$.
} 
\label{fig5}
\end{figure}
 
The complete determination of the evanescent bands in 2D is a very tedious task
due to the large number of states when varying $\Delta_B$. 
Instead of finding the equivalent to Fig.\ \ref{fig3} for the 2D case,
we have followed a different approach to determine the existence of
Majorana edge states. Since the topological phase transition  is signaled by a complete 
vanishing of the complex wave number, we have scanned the value of
 ${\cal F}_{2D}(k=0)$
when varying the Zeeman parameter $\Delta_B$ of the Hamiltonian. The regions 
between nodes 
of 
${\cal F}_{2D}(k=0)$
are the different 
phases, and we have just studied a few selected values of $\Delta_B$ in each region. 
For a 
given $\Delta_B$, we look for a set of evanescent modes in the complex-$k$ plane. 
With these modes
we then find matrix ${\cal M}_{k'k}$ and compute its lower eigenvalues. As discussed 
above, 
each vanishing eigenvalue is associated with a Majorana zero mode. 

In absence of Rashba mixing each transverse mode 
of the quantum wire behaves as an independent
1D system: there is an onset for the appearance of the $n$-th band
Majorana mode
\begin{equation}
\Delta_{B,n}^{(c)}=\sqrt{\Delta_0^2+(\mu-\varepsilon_n)^2}\; ,
\end{equation}
where $\varepsilon_n$ is the $n$-th eigenvalue of the transverse quantum well,
and multiple Majorana modes are possible independently. This physics is
contained in the results shown as a dashed line in Fig.\ 
\ref{fig5}a where, as could be expected, the derivative discontinuity 
${\cal F}_{2D}(k=0)$ has nodes precisely when $\Delta_B=\Delta^{(c)}_{B,n}$.
We have also obtained that, in this case, matrix ${\cal M}$ has $n$ different zero 
eigenvalues when evaluated for a value of $\Delta_B$ such that
$\Delta^{(c)}_{B,n}<\Delta_B<\Delta^{(c)}_{B,n+1}$. In practice,  
a fast numerical convergence of these lower eigenvalues to $\approx 10^{-9}$ 
is easily obtained.

\begin{figure}[t]
\centerline{
\epsfig{file=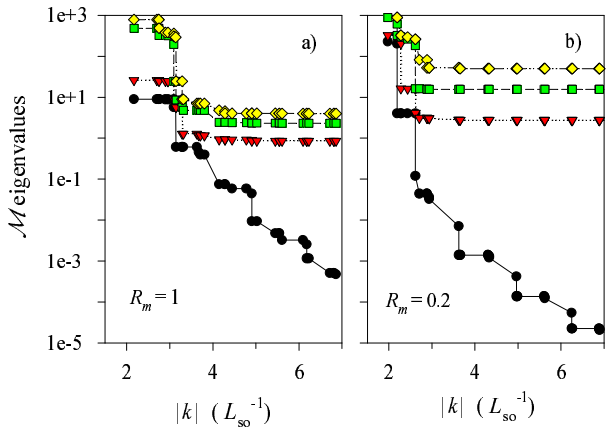,width=0.45\textwidth,clip}
}
\caption{(Color online)
Convergence of the lower eigenvalues of matrix ${\cal M}$ 
when $\Delta_B$ lies between the third and fourth nodes of ${\cal F}_{2D}(k=0)$.
Left panel corresponds to full Rashba mixing ($R_m=1$) with $\Delta_B=4.5E_{SO}$
(cf.\ Fig.\ \ref{fig5}a) while right panel is for an artificially reduced mixing ($R_m=0.2$)
with $\Delta_B=2.5E_{SO}$. The remaining parameters are the same of Fig.\ \ref{fig5}.} 
\label{fig6}
\end{figure}

Conspicuous modifications to the above scenario are seen when 
including the Rashba mixing. Notice that in Fig.\ \ref{fig5} we are assuming a regime
of strong spin-orbit in which 
$E_{SO}\approx\varepsilon_{n+1}-\varepsilon_n$.
We find that for $\Delta_B<0.71 E_{SO}$, below the first node of Fig.\ \ref{fig5}a 
solid line, the matrix ${\cal M}$ has no zero eigenvalue. 
In the region between the 
first and second nodes, $0.71 E_{SO}<\Delta_B<2.17 E_{SO}$, we find
clear evidence of a zero eigenvalue, as proved in Fig.\ \ref{fig5}b. 
Convergence is, however, rather slow and a large number of evanescent
states is required. As a practical scheme, we have included the evanescent states
in an ordered way, taking the modulus of the complex wave number as ordering
parameter. This criterion gives, for instance, that in Fig.\ \ref{fig5}b
for $|k|<7 L_{SO}^{-1}$ we include a total of 44 evanescent modes.

In the region $2.17 E_{SO}<\Delta_B<3.97 E_{SO}$,  between the second and third
nodes of Fig.\ \ref{fig5}a solid line, the lower eigenvalues of matrix
${\cal M}$  converge to finite values, as shown in Fig.\ \ref{fig5}c
for $\Delta_B=2.5E_{SO}$. This behavior is due to the Rashba mixing, 
for, in its absence, we find two zero eigenvalues in this region. Though
numerical, the difference between panels b) and c) of Fig.\ \ref{fig5}
is quite clear and leads to the conclusion that Rashba mixing 
hinders the coexistence of two Majorana modes, which is in 
agreement with the 
results of Ref.\ [\onlinecite{lim12}] for finite systems. 
It is remarkable that
this hindrance persists even in the semi-infinite system.

We focus next on the existence of edge states when $\Delta_B$ lies 
between the third and fourth nodes of ${\cal F}_{2D}(k=0)$; that is, when 
three zero modes are present without Rashba mixing.
As shown in Fig.\ \ref{fig6}a, one Majorana mode again 
emerges when enough evanescent states are included  in matrix ${\cal M}$.
This result agrees with the physical scenario found in 
Ref.\ \onlinecite{pot10} using a tight-binding approach; i.e., 
just a single zero mode is protected in non trivial topological phases.
For the sake of discussion, we have artificially weighted the Rashba mixing
term by a factor $R_m$ in Fig.\ \ref{fig6}. Convergence with the $|k|$ cutoff
is much slower with full mixing ($R_m=1$) than with a reduced one ($R_m=0.2$),
clearly indicating the relevance of this mechanism.

\begin{figure}[t]
\centerline{
\epsfig{file=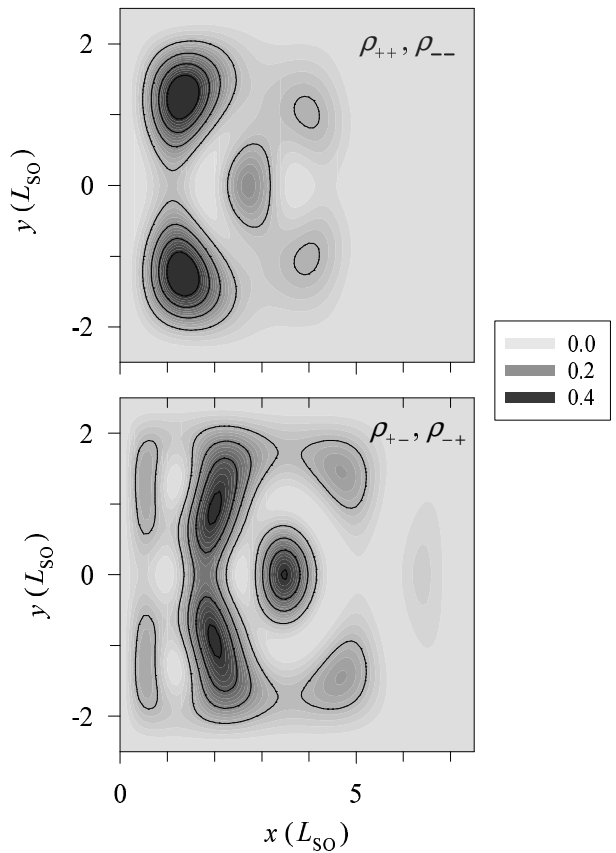,width=0.45\textwidth,clip}
}
\caption{
Density distributions for the Majorana zero mode 
shown in Fig.\ \ref{fig5}b. Arbitrary units are used
in a gray color scale. Contour lines are displayed for a better
presentation. 
} 
\label{fig7}
\end{figure}

For completeness, Fig.\ \ref{fig7} shows the density distributions
corresponding to the Majorana mode of Fig.\ \ref{fig5}b,
using a 2D generalization of Eq.\ (\ref{eq14n}). Notice that, on the
scale of the figure, the density vanishes for $x=0$ and all values of $y$ which is 
indicating the good convergence of the boundary condition with the 
number of evanescent modes. Of course, the densities also vanish 
asymptotically for $x\to\infty$ and,
as in the 1D case, remain invariant when inverting spin and isospin.

\section{Conclusions}

We have discussed the physics of Majorana modes in semi-infinite
1D and 2D wires using the complex band structure approach. This formalism
provides a natural description of the zero-energy edge modes as
superpositions of multiple evanescent waves. The boundary condition
of a vanishing wave function at the edge cannot always be fulfilled,
this limitation defining the parameter regions (phases) where Majorana modes 
exist. 

The phase transition occurs when for zero energy the
complex wave number of an evanescent band vanishes. 
The phase transition is seen as the emergence of a zero
eigenvalue of matrix ${\cal M}$, defined in Eq.\ (\ref{eq23}). 
In 1D the analysis with 
matrix ${\cal M}$ is equivalent to
the determinant of state amplitudes; while in 2D matrix ${\cal M}$ 
allows to 
monitor the convergence of the lower eigenvalues when the number
of evanescent modes is increased.
In 2D we find evidence that, in case of a strong Rashba coupling, 
the Rashba mixing effect hinders the coexistence of multiple
Majorana modes in the semi-infinite wire. In agreement with tight-binding
models,\cite{pot10} 
alternating regions having zero and one Majorana modes are found when 
increasing $\Delta_B$.

\begin{acknowledgments}
This work was funded by MINECO-Spain (grant FIS2011-23526),
CAIB-Spain (Conselleria d'Educaci\'o, Cultura i Universitats) and 
FEDER. Discussions with J. S. Lim, R. L\'opez and D. S\'anchez  are 
gratefully acknowledged.
 
\end{acknowledgments}


\begin{thebibliography}{44}
\bibitem{note} 
In 1D the analytic form of
the Hamiltonian energy eigenvalue as a function of $k$ is known (see, e.g., Ref.\ [\onlinecite{ore11}]). 
Therefore,
determining the allowed $k$'s amounts to finding the (complex) roots of an eigth degree polynomial
in this case. However, the extension to 2D of this approach is, at the very least, unclear since the 
analytic eigenvalues are unknown in 2D.
\end{thebibliography}
\end{document}